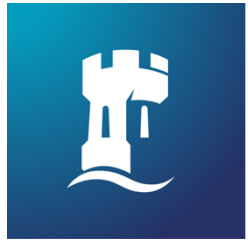

# Colouring the sculpture through corresponding area from 2D to 3D with augmented reality.

Submitted September 2020, in partial fulfilment of the conditions of the award of the degree **MSc in Human-Computer Interaction**.

Yaozhong Zhang

Student ID: 20145930

School of Computer Science

University of Nottingham

I hereby declare that this dissertation is all my own work, except as indicated in the text: Signature ______________________

Date 10/09/2020

I hereby declare that I have all necessary rights and consents to publicly distribute this dissertation via the University of Nottingham's e-dissertation archive.

# Abstract

With the development of 3D modelling techniques and AR techniques, the traditional methods of establishing 2D to 3D relation is no longer sufficient to meet the demand for complex models and rapid relation building.

This dissertation presents a prototype development implemented that creating many-to-many correspondences by marking image and 3D model regions that can be used to colouring 3D model by colouring image for the end user. After comparing the three methods in the conceptual design, I chose the creating render textures relation to further development by changing to Zeus bust model and connecting the AR environment. The results of testing each part of the prototype shows the viability of the creating render textures relation method. The advantages of it are easy to build many-to-many relations and adaptable for any model with properly UV mapping. But there are still three main limitations and future work will focus on solving them, building a database to store relation information data and printing 3D coloured models in real world.

# Acknowledgments

Thank you to the faculties and staffs at the University of Nottingham for their continued enthusiasm for work during this difficult time.

Thank you for Associate Professor Paul Tennent, your guidance is really helpful on my project. You give me more inspiration when we discuss the project during every online meeting. Thank you for your work, I really enjoyed discussing with you.

Finally, I would like to thank my mother and father for their support for my studies. Thank you, my families, friends for their care and concern during my studies.

# Table of Contents

# List of Figures

# List of Tables

# 1. Introduction

## 1.1 Research Background

Augmented reality colouring technology creates a more vivid experience for the user by combining it with traditional colouring methods [1,2,3]. The most appealing aspect of AR colouring is that it is not limited by space or locations, which means that in today's world of smart devices, the user can access the program in any scenario by using any set object as a target while using the tools within the program to complete the colouring operation.

Many sculptures have faded due to historical or human factors, and with the help of modern science, archaeologists and scientists have discovered the colours of some ancient sculptures to unveil their original appearance [4,5,6]. In 2008, the reconstructions made by archaeologist Vinzenz Brinkmann and Ulrike Koch-Brinkmann in exhibition *Gods in Colour* showed to viewers many vibrant pictures of the former polychrome of the sculptures [7]. Normally, experts often use artificial repair work and 3D modelling software, such as *Maya* or *Blender* to restore or recreate sculptures model. Acropolis Museum use digital superimposition of original colours onto a 3D computer generated image of the *Peplos Kore* sculpture [8] and also create 2D online digital interactive game *Colour the Peplos Kore* for visitors using colours and brush they choice [9]. Using digitization tools to quickly help scientists and archaeologists construct and repair polychrome sculptures not only saves time and resources, but also can provide visitors with an educational and fun experience.

However, some of the exist methods must reduce the impact due to the volume and surface complexity of the sculpture by reducing the volume or ignoring details, which would compromise the physical characteristics of the original sculpture [10]. Majority of the AR colouring work so far has been based on textbooks or pages, focusing on hand-drawn textures for display on 3D models [1,2,11], and some of the work is based on direct colouring to 3D models, interacting with models manipulating in virtual space [12,13]. Therefore, the establishment of 2D to 3D correspondence relation through AR technology still has broad application potential.

## 1.2 Research Gap and Motivation

After investigating and experimenting with currently viable methods of model colouring and comparing them, first of all I found that they all have strengths and weaknesses that are difficult to compensate for each other, they all directly satisfy the users with colouring needs using the matching established by program's producer [1,2,3,10,11]. In addition, the models are also relatively simple, which may be related to the amount of workload and ease degree of modelling of the program, as well as the final commercial presentation [28,30].

As the technical barriers to modelling become lower and lower, and a large number of models are now available for free download online, the fastest way to improve efficiency is to build a program that allows users to create their own correspondence relations between models and images, the relations also could be used by others, so that more models could be developed and used. It would directly impact the museum's work to restore the polychrome sculpture.

## 1.3 Project Aim and Question

With the aim of allowing users to quickly obtain the 2D to 3D correspondence and use it to colour the 3D model through image with AR, this dissertation describes a prototype design that can colour 3D sculptures through images. In the prototype, users can use a mobile device with a camera to scan the image target into the program interface and use a haptic screen and capacitive pen to create a matching area between the model and the image. By identifying the location of a finger or capacitive pen on the image, the user can colour the area according to their preference or reference colour. The prototype can be used to colour pages with different image target in textbooks, electronic devices in the home or museum environment for helping restoration, inspiring artists, and providing education for the public and children.

Research question: How to implement colouring the sculpture through corresponding area from 2D to 3D with augmented reality technology?

## 1.4 Dissertation Structure

The main content of the dissertation begins with the review of three-dimensions paint and augmented reality colouring works currently applied in Chapter 2. A particular attention is paid to the method of establishing corresponding regions between 2D and 3D, which helps to properly conduct this study in the context of current research with a broad perspective. In order to achieve the project aim, I have developed a prototype in Chapter 3. In the conceptual design, I have used three methods to determine the implementation flow and model applicability with using different models for comparison. In the detail design, I have chosen an appropriate method to connecting with AR interface. Chapter 4 documents the content and results of the prototype test on the iOS platform. The following Chapter 5 then discuss the interpretations and implications, the limitations with existing prototype and recommendations for future work. Finally, Chapter 6 provides a brief conclusion of the entire project and dissertation.

The main contributions are:

1   Implementation of conceptual design using different correspondence methods;
2   Comparison of the strengths and weaknesses of different methods and the types of models to be applied;
3   A prototype for creator and end user is implemented;
4   Constructing a colouring system in the AR world.

# 2.　Related Work

## 2.1 Three-Dimensions Paint

In three-dimensional computer graphics, applying texture mapping is a common way to increase the apparent detail of an object without increasing the complexity and details of the model [14]. In the last three decades, the development of 3D painting has been pursuing a way to change the texture map of 3D models using brushes and colours simulated by computer technology in virtual space. Developers realized that textures have a very important influence on the appearance of the model, so much of the basis for 3D painting is also based on textures.

In1990, Hanrahan and Haeberli introduced an interactive paint program，which allows users to paint different types of paints and materials directly onto 3D shapes [12]. Users can utilise the tablet to control the position of brushes containing shapes and paints, and the created material properties are stored as a set of associated texture maps in the end. The main contribution of this work is using rendering techniques as the user paints to create the appearance of the surface of object and the results of the interactive colouring is a reproduction of the 3D shape. The author also gives a full description of the problems of creating and mapping texture maps, firstly the problem of defining the coordinates and orientation of textures, and secondly the problem of deformation of texture maps due to parameterization. Subsequent works in 3D painting has also evolved in the direction of a more realistic view.

Distinct from traditional 3D painting programs that use predefined UV maps, Takeo Igarashi and Dennis Cosgrove present a technique for creating effective UV maps for texture painting programs, this system of systems dynamically creates tailored UV maps for newly drawn polygons during the painting process. The user can draw undeformed strokes over the entire surface from any direction. Dynamic texture distribution allows the user to draw smooth strokes at any zoom level. This technique can be effectively implemented using standard 3D rendering capabilities without the need to write code for pixel-level manipulation [15]. The result of the drawing is stored as a standard texture polygon model and can be read by various graphics applications. This work get inspiration for texture generation and preservation for 3D painting. Lu et.al. allows the user to generate texture maps by drawing directly on the surface of the 3D model using a defined UV map in relation [16]. During the painting process, when the user paints strokes in the 3D view of the object, the system subsequently re-projects the user's strokes to the corresponding positions in the 2D texture bitmap based on predefined UV mappings (texture coordinates).

The proliferation of touch screens has made 3D painting no longer limited to computer programs. Many previous works have demonstrated that using a haptic device to draw textures directly from the surface of an object is an effective technique for creating textured images intuitively and interactively [17,18,19,20,21]. In 2008, Wakita et.al propose a system for designing haptic models based on textures for haptic devices. In the system [22], differences in tactile impressions are rendered by dynamically varying the magnitude and/or direction of the reaction force based on the pixel values of the object surface, which maps the particular texture image and converts roughness, stiffness, and friction into a 2D image.

With the advancement of technology, the development of software has made it more realistic for 3D painting to be done. One of the most important developments is the render texture. Render textures are special types of textures that are created and updated by specifying a camera at run time in Unity software [23]. The Camera component has a *Target Texture* variable, which can be set to a *Render Texture* object to output the camera's view to the texture rather than the screen [24]. Monitoring the display is a good metaphor for how it works: the camera is the camera, the rendered textures are the video recording, and the material is the screen your video is projected onto.

In 2015, the *RealTime Painting* project [25] published in Unity assess by Rodrigo Fernández. The developer using render textures to simulate painting the texture of material of the mesh in real time. In this project, the camera of the render texture looks at a quad with the UV map of the model. And then anything that is placed in front of the camera will be added the texture. Painting mesh in this way will allow us to have a lot of freedom and will have very clean results for the end users, for now, after merging texture it won't allow users to roll back the changes made.

## 2.2 Augmented Reality Colouring

The definition of augmented reality (AR) used in this project refer the survey proposed by Ronald Azuma in 1997 and a book written by Alain Craig in 2013. The survey defines AR as any system that has the following three characteristics: combines real and virtual; interactive in real time; registered in 3D [26]. In 2013, Alain Craig further extends the augmented reality definition, not simply including the essential components of AR, but also defining from users' perspective. He defined augmented reality is a medium that user's experience that is interactive, at least by changing perspective dependent on location and perspective of the viewer and the digital information is overlaid on the physical world, that is in both spatial and temporal registration with the physical world and that is interactive in real-time [27].

The target users of the majority of AR colouring works are children and based on the context of colouring books or colouring page. In early AR colouring works, it common to use image registration and image inpainting techniques to map or extract the image texture to the visible and occluded regions of the 3D model. The users could utilise the smart devices with camera scanning and recognizing the colours and brush strokes to forming the model wrapped texture map on screen. Clark and Dunser at The HIT Lab NZ in 2012 have created an interactive AR colouring book that automatically performs registration and texture extraction from colouring book pages. Allowing the user to innovate the colour of the book's content and automatically mapping it to virtual pop-up 3D scenes and models based on the user draws with a coloured pen. The key to the technique is image registration of pages that has been colour removal by computing the RGB colour threshold within a pixel through natural feature registration. The software identifies the black and white line art as the baseline which by the light intensity and the saturation of colour on the drawing [1]. Currently, *Quiver* [28], a company developed by The HIT Lab NZ, is producing and releasing commercial products about AR colouring.

Lee and Choi use Vuforia SDK (Software Development Kit) with Unity 3D to bring more possibilities to AR colouring book. They propose an algorithm to extract a specific rectangular region from the live video and use it for colouring book's texture map. The work involves the processing of image deformations by calculating the transformation matrix from local coordinates to world coordinates, and from world coordinates to screen coordinates. In addition, the perspective created in reality due to the deviation of the camera from the page angle was corrected [29]. Basing on the key technology, the AR colouring puzzle, *Dinosaur Sketchbook* [30], was built and then was developed commercially, which provide its technical viability [3]. After that, Cho et al. used colour image acquisition and processing algorithms, and matrix computation and image processing algorithms to build a colouring game-*Playing*

*House.* The aim is to establish connections between the real world, the AR world, and the VR (virtual reality) world, so that all worlds are seamlessly connected together, enabling users to access the same content in different visual ways [31].

Norraji and Sunar built wARna (Wonderful Augmented Reality and Arts), a mobile-based interactive augmented reality colouring book, by texture extraction and mapping using benchmarking and image processing techniques, the colouring system extracts textures within the frame and maps them to the corresponding 3D model by detecting the marker frame [11]. After that, they developed colour recognition and colouring programs for preschool learning based on wARna project as educational purpose [32]. This work uses printed line arts provided by the manufacturers, thus avoiding the problem of curved pages caused by bookbinding.

A major limitation in these works is that users cannot fully create 3D content. While users can colour 3D content, they will always need to prepare publisher-supplied print art lines. In order to increase interactivity and address the cause of user boredom caused by the post-colouring model's view-only functionality, many colouring works are also actively exploring the use of post-colouring models.

To the problem of camera's angle deviation from the page arising in the using of the system, Magnenat et al. present a real-time variable surface tracking and texture synthesis method for the AR colouring process. Deformable surface tracking allows warped painted pages to be tracked in real time on mobile devices, increasing the interactivity of the experience [2]. Real-time texture compositing uses textures drawn by the user on the 2D image and applied in real time to solve mirroring and parameterization seams artefacts, which proposed algorithm based on both diffusion-based and patch-based methods, as it both extends the known parts of the image to the unknown ones, and copies pixels know regions to unknown ones. According to the testing results, texture composites aren't perfect, but acceptable to most users. Unlike other AR colouring, which can just see the coloured model after colouring, live colouring is more interactive for the user in real-time. MagicToon solves the problem that AR content cannot be freely created by users, the system can automatically build 3D models according to user-drawn 2D images and interactively edit more complex scenes in AR screen [10]. The system uses NaturaSketch [33] is a diffusion-based and patch-based inflation method that encodes each pixel by calculating its distance to the nearest black pixel and extends the drawing to the entire area using the inflation method. This method can be used for symmetrical 3D models, with the limitation that the original texture on the image is used as a texture for the front and back surfaces of the 3D model.

From the above summary of works, most of the works are based on changing or generating texture maps for implementing colour modified. For this project, redesigning the user interface is important when developing mobile AR colouring prototype. Moreover, establishing one-to-multipoint mapping relations requires consideration of finding ways from the model generation steps.

# 3. Prototype Development

This project is going to use prototype development method to try a technical task of colouring sculpture, the focus is on the feasibility of the idea to evaluate its performance or the performance of some of its components. In the iterative process of the project, it is easier to make rapid prototyping in the conceptual design and compare the complexity and usability in each concept prototype and add interactive components that are missing in the process of implementing functions. The designer uses this prototype to consider risks and to determine which components are needed in the final version. This can include testing the user interface or evaluating specific technical components to find out what is suitable for the users.

The crux of this work is in building a correspondence relation between a front view image of 3D model and a texture map wrapped on 3D model. Ideally, this relation can be established by different users, then the end users could utilise it to transform the colour of 3D model by colouring on the image.

As a result of current COVID-19 lockdowns, I am not going to attempt to user study and the project will focus instead on the technical challenges. The project is developed by Unity 3D and the AR part is designed using Vuforia, the detailed information will be presented in the next section.

## 3.1 Background

### 3.1.1 Unity and Vuforia

Unity is a cross-platform game engine that provides interactive real-time 3D content creation and supports C# language scripting developed by Unity Technologies from 2005. This project uses Unity 2019.2.8 f1 version as the development environment.

**Figure 1 Unity interface [34]**

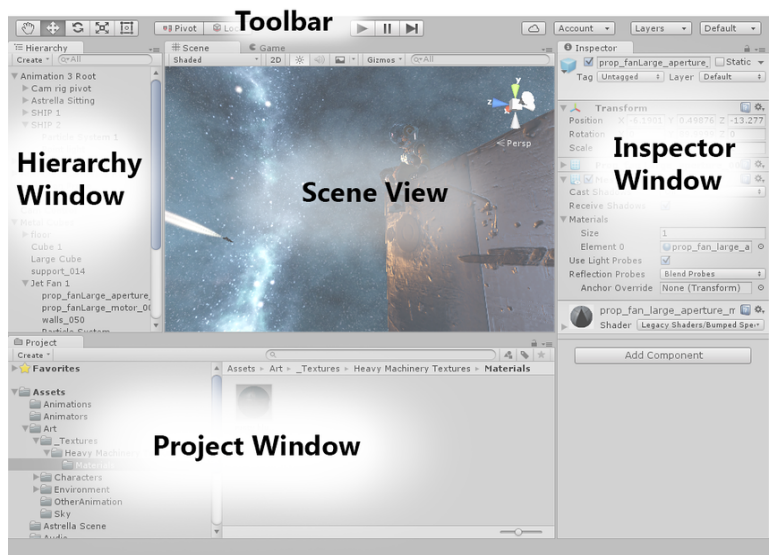

The *Toolbar* provides the most basic functionality for working with objects, including the basic tools for manipulating the scene view and the objects in it on the left and the play, pause, and step controls in the middle. The *Hierarchy Window* shows the texts names and hierarchical representation of each object in the *Scene window*. The user can view and edit all the properties of the currently selected object in the *Inspector Window*. The *Project Window* displays all the assets that are available to use in the project (Figure 1). For the more detailed description, the renders can check out the Unity manual [34].

This project utilises the Vuforlia version 8.5.9. As of 2017.2 version, Unity integrates the Vuforia Engine, which makes it easier to create the most advanced augmented reality experience for handheld and head-mounted display (HMD) devices. The advantages of Vuforia are multi-platform release and 3D object target. The project can be load on iOS, Windows and Android system, which provide convenience for subsequent expansion of the system and full platform coverage. Moreover, 3D object target might provide a realistic way for users triggering system and viewing completed AR works in the real world. Reader can learn more about the Vuforia Engine platform at Vuforia developer library [35].

### 3.1.2 Blender

Blender is a cross-platform application and open source 3D creation suite, which can be used to create 3D visualizations such as still images, 3D animations, VFX shots, and video editing. Blender also has relatively small memory and drive requirements compared to other 3D creation suites [36]. This project uses Blender version 2.83.4. Blender's interface is separated into three main parts: *Topbar* at the very top: *Areas* in the middle, *Status Bar* at the bottom (Figure 2).

**Figure 2 Blender's default Screen Layout [36]**

*Topbar* (blue), *Areas* (green) and *Status Bar* (red)

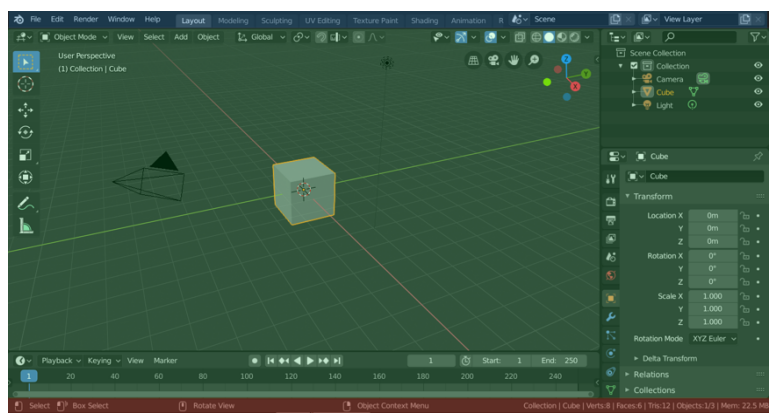

### 3.1.3 Three-Dimensions Model

This section will introduce the 3D modelling process and the models used in the prototype development (Figure 3).

The normal process for 3D modelling is firstly to build a low poly model in the modelling software including the skeleton of the model and the large-scale region. The second step is to enrich the details on the basis of the low poly model by increasing more structures and the number of vertices, edges, and faces, then establishing a high poly model. The third step is to topology the high poly model to the low poly model for reducing the number of faces on the high poly model to ensure the system with high speed and quality for model generation. The fourth step is to UV unwrapping. The low poly model needs an unwrapped UV map to define how the texture map wrap on the model. The fifth step is texture baking. The process is to map the detail information including normal map with bump information, colour map, metallic map, roughness map on the high poly model to the low poly model. The final step is to adjust the parameters for getting ideal model.

**Figure 3 Three-Dimensions modelling diagram**

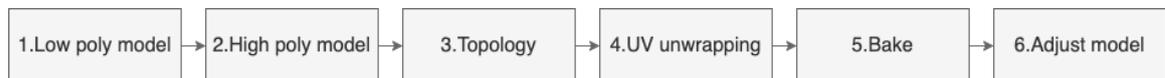

From the modelling process, it can be analysed that the following resources could be used to transform the colours: model, UV map, and texture map.

In the next chapter Conceptual Design, these resources will be used to try to develop a prototype with comparing different methods. The following introduces the models used in the development process.

### 3.1.3.1  Clossal Marble Bust of Zeus

For the 3D model, I referenced sculptures from the British Museum, many of them are now digitally conserved, including the creation of digital models of the statues for public study. As the statues themselves have been damaged for historical reasons, the fragmentation of the models can affect the effect of the colouring. So, I chose to base my work on a more complete bust of Zeus digital model (Figure 4). The bust is a composite object with the head being antique roman, the nose, body, and socket base are 18th-century restoration made to fit the head, which was probably an over-life size statue originally [37]. The model updated by Thomas Flynn in 2016 by *Agisoft PhotoScan* using 171 photos [38]. The other reason of choosing this 3D model is its UV mapping properly, and in texture mapping it is difficult to get the model position by pixels using the method in the above work. This is challenging for the next prototype development.

**Figure 4 Colossal marble bust of Zeus model [38]**

3D model (top left) and texture map (top right)

3D UV map (bottom left) and 2D UV map (bottom right)

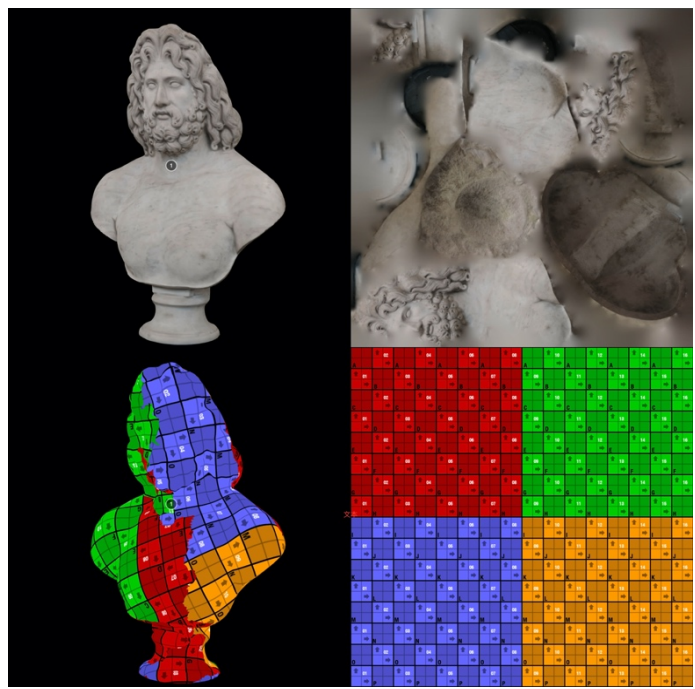

### 3.1.3.2  Caligula Digital Sculpture

In 2010, the National Endowment for the Humanities grant was given to the University of Virginia and Virginia Museum of Fine Arts to undertake new studies on Caligula (Figure 5). The aim of the technical studies was to create a digitally restored statue model and to study its ancient polychrome. By scanning the marble statue and using digital technology to repair the damaged surface, missing arms and hands of the marble statue, a painted digital model was finally created. The colour added on the digital reproduction model is part of the ancient marble statue [39]. The restored Caligula statue in coloured toga has been generated as an interactive model for people to learn and browse online. For more detailed information, readers can check the Virtual World Heritage Laboratory website about introduction of the Caligula digital sculpture project [40]. This model is chosen because it is a restored model, and it has no texture, and the UV mapping is not regularly wrapping on model.

**Figure 5 Caligula Digital Sculpture in white tunica and purple toga [40]**

3D model (left) and 3D UV map (right)

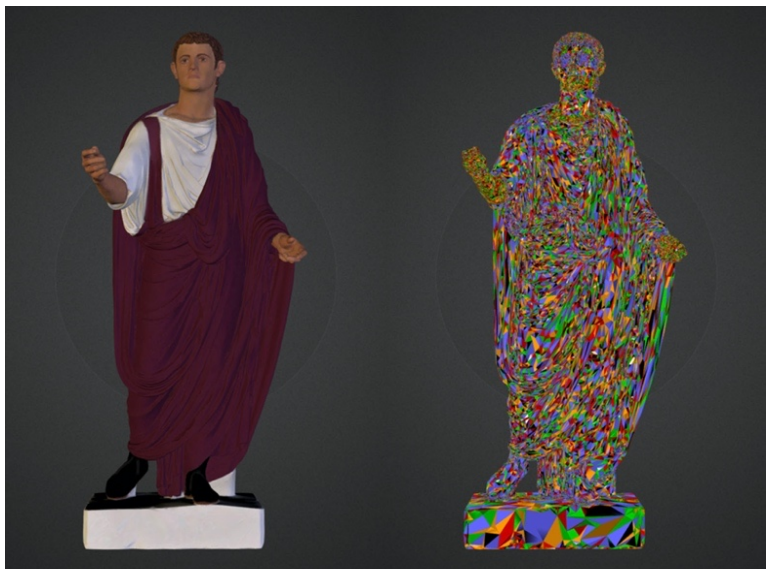

## 3.2 Target User Group

The target group of users for this project includes museum colour restorers, as well as members of the public with colouring needs. The end users can take advantage of a relation that can be created in a relatively simple way by the creator.

Most of the works previously described in Chapter 2 were designed to face the user directly for colouring, with the developer specifying the 2D to 3D correspondence, and the user using this relation to colour the model, generating colours or brushstrokes. But with the increasing simplicity of 3D modelling technology, such as *RealityCapture* [44] software, that only needs to take multi-angle photos of a model, there will be more and more resources could be used to develop in the future. Therefore, the workload of developing program by traditional workflow is enormous for companies or individuals.

In this prototype I consider two types of users: the creator, who can use the system to quickly establish correspondence; and the end user, who uses the correspondence to colour sculptures.

## 3.3 Conceptual Design

In the conceptual design, I consider the following three methods to implement 2D to 3D area building relations. This is to identify the steps of different methods to achieve the functionality. After using the simple model to achieve the relations, I evaluated the strengths and weaknesses of methods by using the more complex model presented above as an example.

### 3.3.1 Model Segmentation

As explained modelling process in the 3.1.2, model segmentation can be done in two ways. The one way is to model each component separately and then form a complete model. The other way is to split the UV mapping of the model according to the demand and then controlling colour of each component by changing material on it. Both 2D images consist of the corresponding area of each 3D part.

To achieve the above purpose, I built a 'snowman' model by stacking two spheres, the head is smaller than the body, and the 2D image is made up of two different size circular sprites for corresponding to the 3D model. The system needs to implement 3D head and body colour can be controlled by 2D correspondence area.

As figure 6 in Unity software, the colour of head and body of snowman model has been changed into red and black by selecting the right area of the corresponding sprite and then selecting the fill colour by the left colour selector.

**Figure 6 Before filling (left) and after filling (right)**

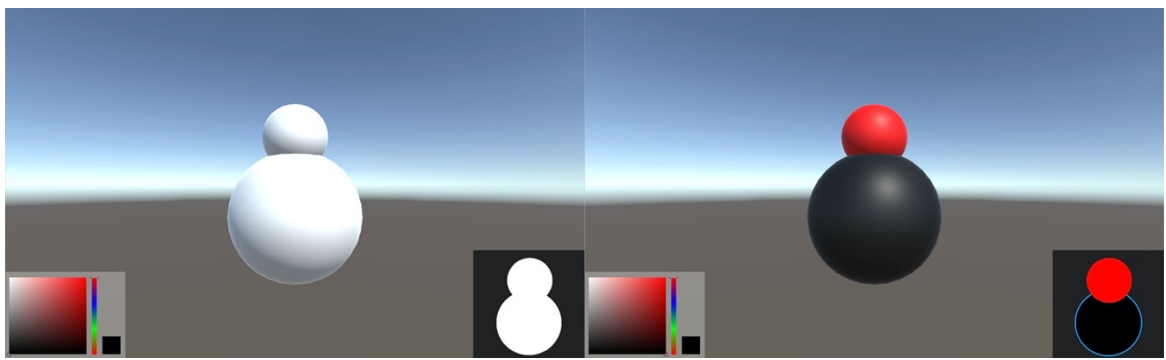

Since the construction of the snowman model is very simple, and there is not much difference between the two ways in the final effect, so in order to further verify the difference in the effect of the two methods, I chose the more complex Caligula model.

I topology the Caligula model, perform UV unwrapping, and then divide the part of the model into 9 groups (Table 1)

**Table 1 Groups of Caligula model regions**

| Group No. | Region |
|---|---|
| 1 | Hair and eyebrows |
| 2 | Eyes |
| 3 | Pupil |
| 4 | Lips |
| 5 | All skins |
| 6 | Tunica |
| 7 | Toga |
| 8 | Shoes |
| 9 | Statue base |

Considering the limitations of the current hardware and software, the UV unwrapping of the model can only be done after the topology enters low poly model. After the topology, the high poly model has become a low poly model, and the details of the face region cannot be restored, so eyes and pupil have to be put into together as one group. According to the UV mapping, the model could be controlled by 8 materials in Unity software in total. The final effect is as figure 7.

**Figure 7 Coloured Caligula low poly model in Unity**

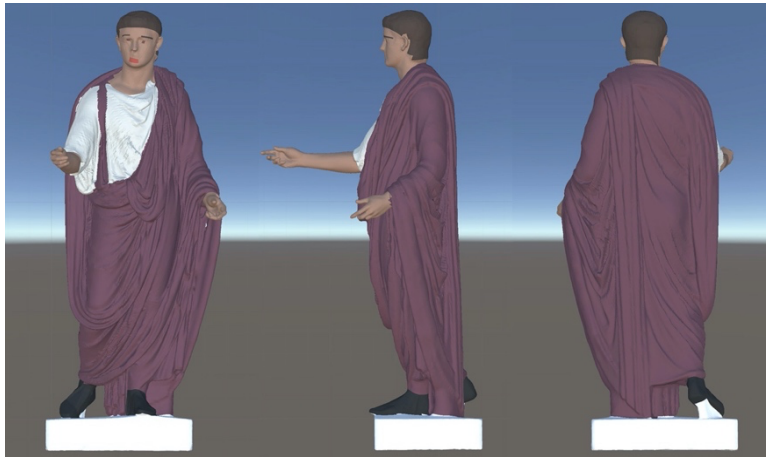

In splitting way, there is no need to consider about modelling, but topology and then UV unwrapping is still required. Due to the low poly model, the facial features are damaged very seriously, so that the appearance characteristics can no longer be distinguished. It lacks the details of high poly model and also cannot use bake texture of high poly model; the hair and clothes folds on the model have also lost their original structure; the junction between each component is also severely jagged due to the shape of faces on the model; the final effect can only restore the outer shape of the original model but not details. The result shows this approach can only achieve acceptable results with simple geometry model.

In order to achieve acceptable effect on more complex model, I began to try remodelling each part and then combine them together, so that the details of the model can be restored using high poly models. The figure 8 shows the effect of colouring high poly model. In this time, the model is divided into 9 groups according to the previous expectation and high poly model adds more details. After remodelling separately, the details can be restored. As result of which complexity of the eye structure, more details of the eye, such as sclera, eyelashes, are still not displayed through the model structure; the process of remodelling and combination of the model takes long than expected; inevitable seams generated in the modelling and combination process; the designed 2D to 3D relations cannot be changed.

**Figure 8 Coloured Caligula high poly model in Unity**

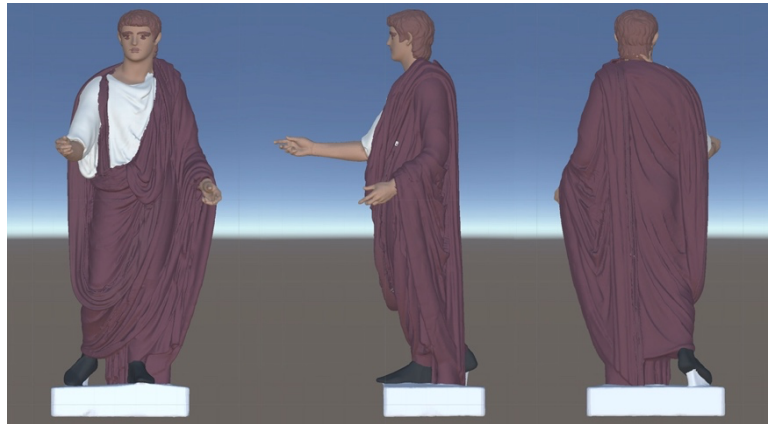

### 3.3.2 Add Multiple Materials with Blender Software

Creator could use open resources software Blender define separate materials on a single 3D object. In *Edit Mode,* dividing the model by selecting the faces to some parts, then adding and assigning the materials for each part (Figure 9). The model could be changing the colours by controlling material.

**Figure 9 Selecting the faces of model and assigning a material.**

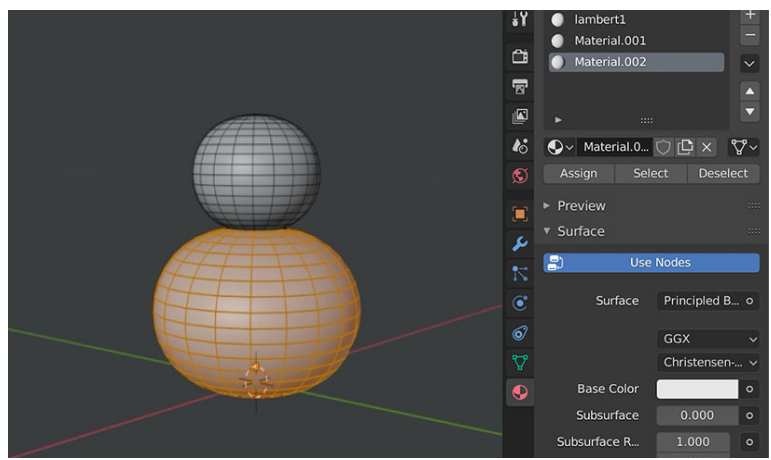

The software is very simple to use and addresses the shortcomings of the previous method. After the implementation on Zeus bust, I found the following features. First of all, this method eliminates the need to build each part of the high poly model separately in order to get a high-quality effect, just utilising a whole model with regular faces, which can exceed the effect of method one, without seams and saving time. Secondly, in this method, the original texture map can be retained after colouring model (Figure 10). It is especially helpful for statues as users seems to be colouring the real collection, which is a more realistic way for colouring.

**Figure 10 Original Zeus bust(left) and coloured skin of Zeus bust (right)**

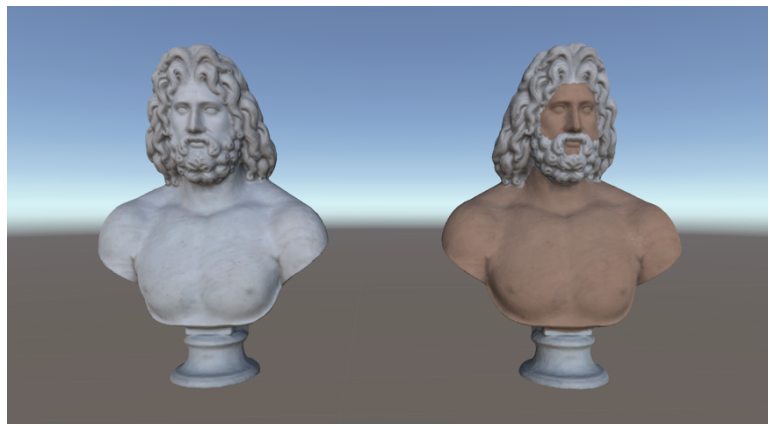

The weaknesses are that this method is still limited by the model. Separating the many faces of a high poly model is a huge amount of work. In particular, the boundary dividing a part of the model is influenced by the shape of the face, and the edges are not always very neat.

Both of these methods require the creator to spend a lot of time dividing up the model sections, but below I consider another method that does not need worry about the model.

## 3.3.3 Creating Render Texture Relation

In this method, I tried to modify the texture map itself, and create a relation between 2D and 3D. Inspired by the aforementioned *RealTime Painting* project, the 3D model can be coloured using render texture in an interactive way, the same can be utilised in 2D image as well. So, my conception is that there are two modes in the system, one for the creator mode and the other for the user mode. In the creator mode, the creator can establish a matching relation between the render texture map and the image by painting the colours and save it locally. In the user mode, the user can change the colour of the 3D model by colouring the front view of the model, using the matching relations that have been established by the creator.

In Unity software, there are three scenes that connect the entire program. *Scene0* is the menu page, which including creator mode and user mode. These two modes are settled in *Scene1* and *Scene2* respectively (Figure 11).

**Figure 11** *Scene0* **(left)**, *Scene1* **(middle)**, **and** *Scene*2**(right) interface in Unity**

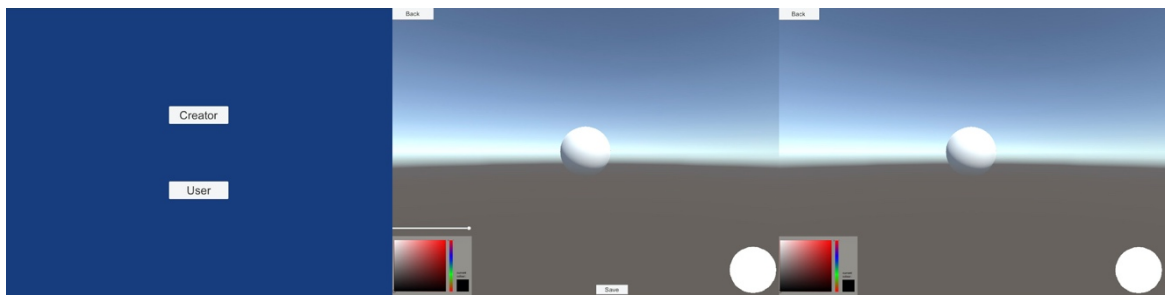

*Scene 0* is a menu interface with *Creator* and *User* buttons in the middle of window, corresponding to creator mode and user mode respectively. In *Scene1,* the interface consists of the following parts: the 3D model of sphere that used in the concept design is showing in the middle of the screen; the colour panel is at the bottom left corner for picking the desired colour by users; the brush control bar is on the top of the colour panel, which is for controlling size of the brush by dragging to the left or right and the system is to use the maximum brush by default; the *Revoke* button that will appear after painting and the *Save* button are both in the lower middle; the front view area of the model, which is circular in the prototype in the lower left; the last one is the *Back* button in the upper left corner, which is used to return to the menu. The layout in *Scene2* is the same as in *Scene1*, except that *Scene2* doesn't need the *Revoke* button, the *Save* button, and the brush size control bar.

**Figure 12 Creator and user flow**

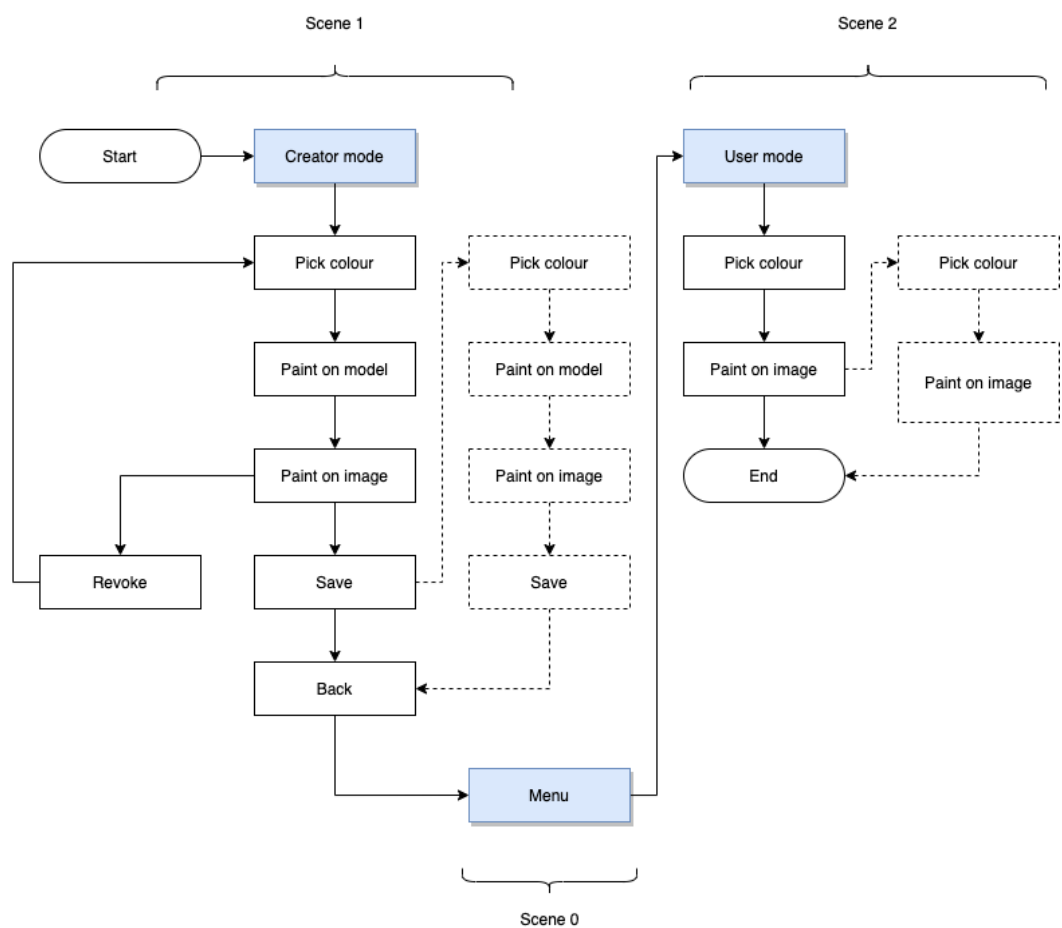

The user flow diagram, as figure 12, shows how the two user groups operate the program. At the beginning, the program will enter the Creator mode by default. In the Creator mode, the creators can select the colour by the colour panel to mark the desired area in the 2D and 3D parts. First, use the brush to paint on the model to mark the area, then performing the same operation on the front view image. In the process of marking the areas. If the creator is not satisfied with one of them, the creator can revoke it. Besides, if one of the parts no marking after clicking Save button each time, the relation cannot be saved, and the system will prompt by missing data. After saving, the creator can choose to continue to create the relation using other colours or return to the menu interface. The user can enter the User mode through the menu interface, they only need to fill in the image by any choosing colour, at the same time the corresponding area of 3D model will be coloured as well.

Technically, in the process of implementing the prototype, the main situations are as follows:

1. The relation between the UV Map and the point of interaction;
2. How to mark on render textures;
3. How to save the relation between render textures for user colouring.

### 3.3.3.1     Pointing at the UV Map:

In order to establish a relation, firstly, the creator needs the marking the mesh. I used the render texture to draw the mesh material texture in real time. The cameras of the render textures of 3D model and image look at materials named *MeshBaseMat* and *UIBaseMat* with the UV map of the model (Figure 13), the only difference between them is the render texture on the image don't need to map to any model. In the prototype, there is no texture on them so that just show a white background for colouring. The materials should have *'Unlit /Texture'* as shader to avoid unusual colouring. So that, anything that is placed in front of our camera will be added the texture.

**Figure 13 *CanvasBase* for 3D model in Unity**

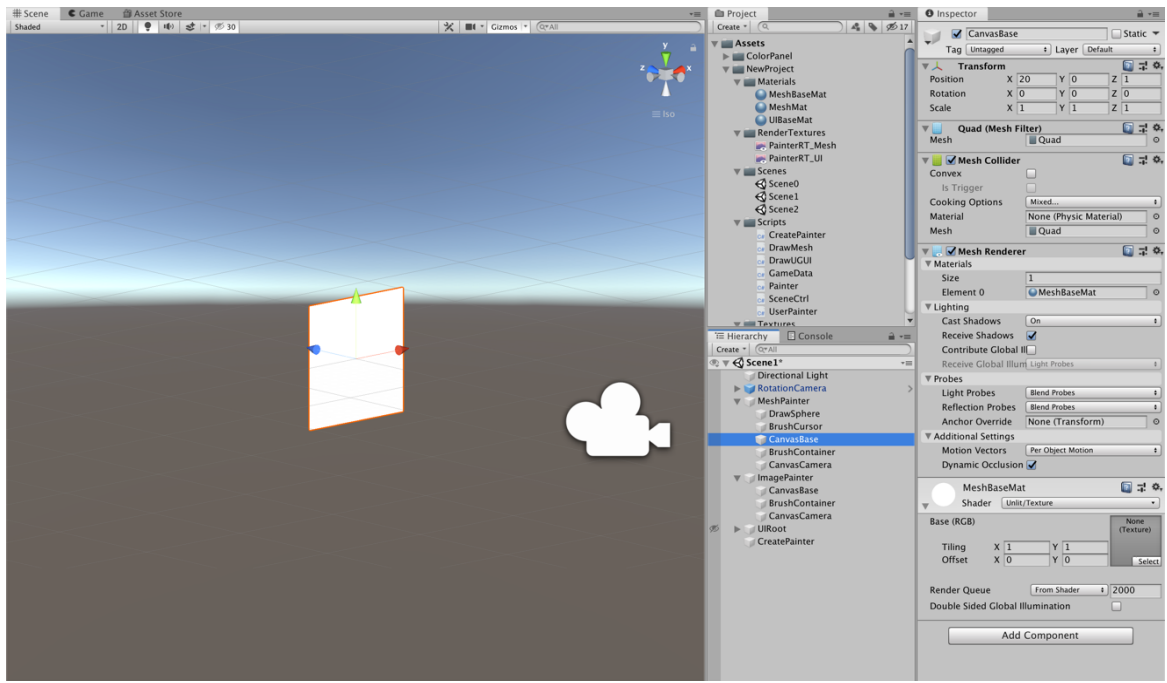

To marking the sphere, I create a material named *MeshMat* using the texture for mesh with '*Standard' Shader* and adding on *DrawSphere* (Figure 14).

**Figure 14 *MeshMat* Inspector**

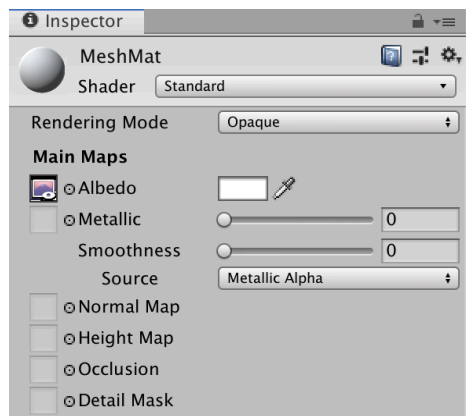

After adding the material to the mesh, I need to use *Raycasting* and a unity function for determining which part of the UV map are users hitting, and then users can place things in front of the camera to simulate colouring over it (Figure 15).

**Figure 15 Code for pointing UV**

```
bool HitTestUVPosition(ref Vector3 uvWorldPosition)
{
    RaycastHit hit;
    Vector3 cursorPos = new Vector3(Input.mousePosition.x, Input.mousePosition.y, 0.0f);
    Ray cursorRay = sceneCamera.ScreenPointToRay(cursorPos);
    if (Physics.Raycast(cursorRay, out hit, 200))
    {
        MeshCollider meshCollider = hit.collider as MeshCollider;
        if (meshCollider == null || meshCollider.sharedMesh == null)
            return false;
        Vector2 pixelUV = new Vector2(hit.textureCoord.x, hit.textureCoord.y);
        uvWorldPosition.x = pixelUV.x - canvasCam.orthographicSize;//To center the UV on X
        uvWorldPosition.y = pixelUV.y - canvasCam.orthographicSize;//To center the UV on Y
        uvWorldPosition.z = 0.0f;
        return true;
    }
    else
    {
        return false;
    }
}
```

The front view will be marked in the same way as above mentioned. In order to display the circle image, I have put it on the texture of the *DrawPanel*, and then all markers will be within the scope of this circle (Figure 16).

**Figure 16 *DrawPanel* Inspector**

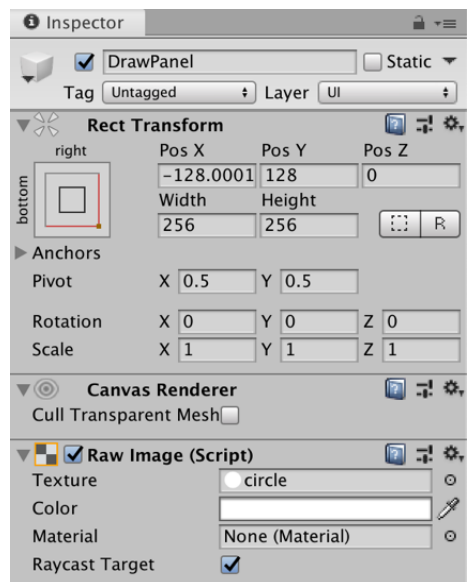

### 3.3.3.2   Marking on Render Texture

To avoid having too many draw calls, I used *Sprites* to change brush instance colour without creating a material instance to allow Unity to group these sprites together.

When the creator starts marking, the *Sprites* will be generated for each created brush under *BrushContainer* (Figure 17).

**Figure 17 Created brush sprites for marking sphere**

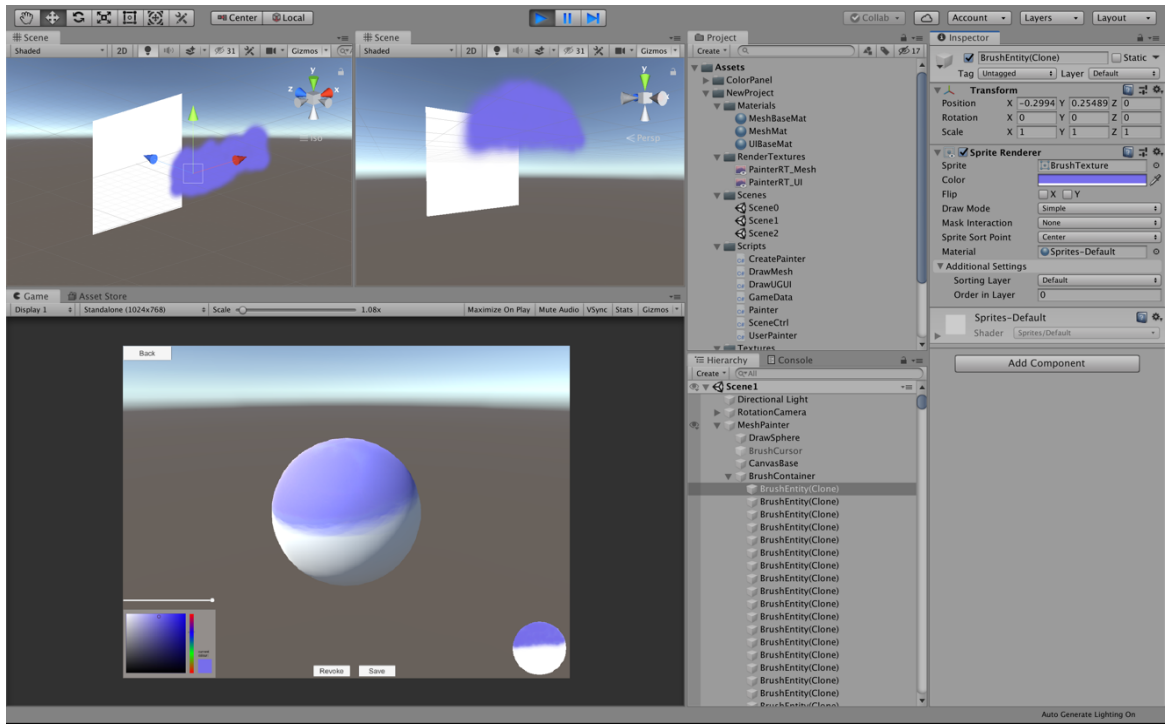

When Sprites are generated, the user can use the *Revoke* button to cancel the

current marking regions by deleting the child object under *BrushContainer* (Figure

18) and then recreate the new relation.

**Figure 18 Code for revoke function**

```
public void Revoke()
{
    foreach (Transform child in brushContainer.transform)
    {
        Destroy(child.gameObject);
    }
}
```

*Save* button is for merging spawned brushes to base textures which are

*MeshBaseMat (instance)* and *UIBaseMat (instance).* After saving, it will clear marks before (Figure 19).

Figure 19 Unsaved texture(left) and saved texture(right)

### 3.3.3.3 Saving the Relation Data

Once the marked parts of the 3D model and image are saved in texture, the coordinate data of those will be temporarily stored in *Game data* (Figure 20). In User mode, the end user will already be able to paint image with these relations.

Figure 20 Code of save data

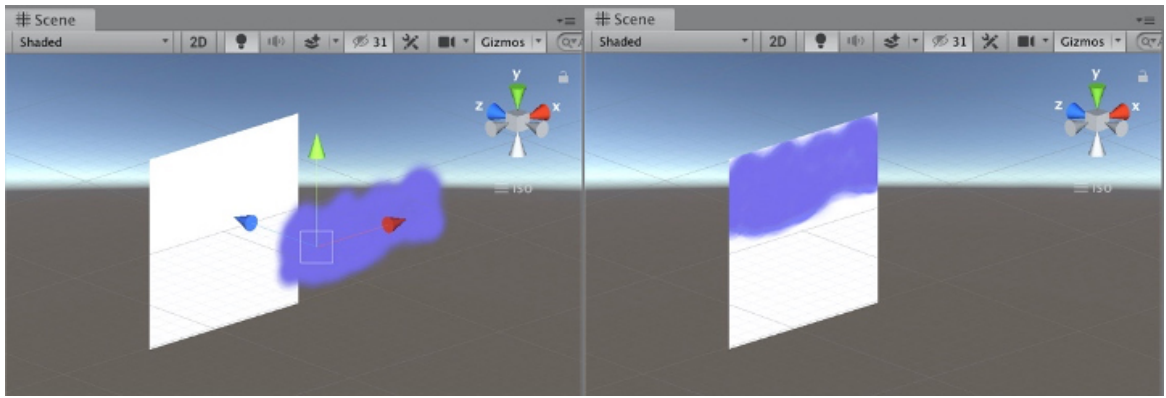

In *Console* window, the it will display ''Missing data'' or corresponding data. the corresponding data are temporarily stored, as shown in the figure 21.

**Figure 21 Debug log for relation data in *Console* window**

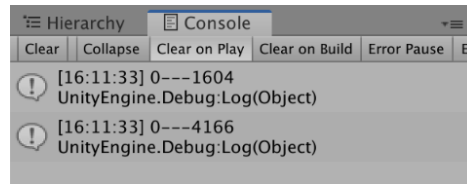

As figure 22, these functions define the coordinates to determine the distance from the current user touch position to the markers coordinate, and if the distance is less than a variable, the corresponding area will be filled with the colour chosen by the end user in a new texture.

**Figure 22 Code of distance detection with brush point**

```
public class TextureData
{
    public List<UIPoint> picPoints=new List<UIPoint>();
    public List<WordPoint> wordPoints=new List<WordPoint>();

    public bool havePoint(Vector2Int _pos)
    {
        foreach (UIPoint x in picPoints)
        {
            if (Vector2.Distance(new Vector2(x.point.x, x.point.y), new Vector2(_pos.x, _pos.y)) < 8f)
            {
                return true;
            }
        }
        return false;
    }
}
```

In the User mode, after the end user selects the colour and fills in the image, the image and 3D model both have new textures with that colour. The figure 23 shows after colour the top of the circle image, the colour of top area of sphere is also be changed.

**Figure 23 Filling red in User mode**

3D texture map (top left), 2D texture map (top right)

The effect in the scene (below)

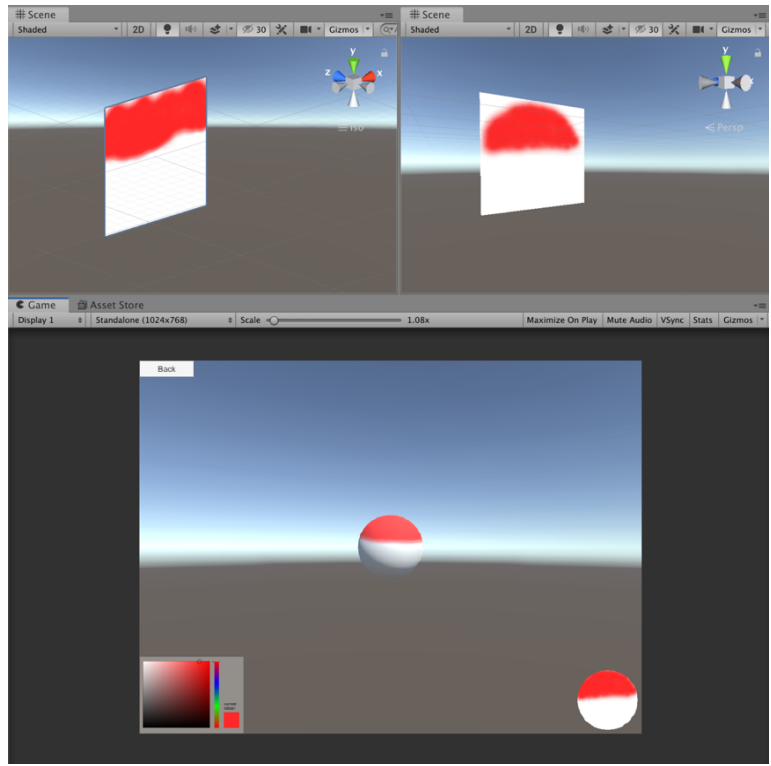

The first advantage of this method is that it can be applied to any model which the UV map is a mesh. It will change the workflow for creating such 2D to 3D model matching as establishing the correspondence in a very simple way for the creator. Secondly, the correspondence is much finer than the traditional method, but it depends on the size of the brush and the distance variable. Thirdly, there is no limit to the size and number of correspondence areas in the relationship building process, it is completely free to be created by the creator. The limitation is still not perfect for fine structures such as eyes, and the user cannot change the colour of an area after filling it. In comparing with the second method, the model cannot use the original texture map as well.

## 3.3.4 Summary

I've compared and summarized three different methods to changing colours of parts of the model. All of them have different advantages and disadvantages (Table 2), and in summary, the third method is more in line with my project aim, so I chose it for detail design and linking AR system.

**Table 2 Comparison between three methods**

| | Method | Description | Advantages | Disadvantages |
|---|---|---|---|---|
| 1 | Model segmentation | Modelling each part | • More details | • Seams<br>• Longer time<br>• Fixed relation |
| | | Splitting the UV | • Model resources | • Topology<br>• Less details<br>• Jagged edge |
| 2 | Add materials | Define multiple materials on a 3D model | • Details<br>• Easier dividing model<br>• Keep original texture | • Too many faces<br>• Uneven edges |
| 3 | Render texture relation | Building relation between render textures of image and 3D model | • Any model with properly UV mapping<br>• Easy to build relation<br>• Mapping accuracy | • Cannot add original texture<br>• Cannot change the colours after filling by user |

| | • Many-to-many mapping |
|---|---|

## 3.4 Detailed Design

With the above summary, I chose the third method for further development. In this part, I mainly introduce about the two parts, one is the setting of the front view and the other AR development process.

### 3.4.1 The Front View Image

In order to achieve the aim of this project, I changed the model in the detail design to a more complex sculptural model, which is the bust of Zeus model with a mesh UV map. The front view in the prototype will need to be turned into a black and white line drawing for created and end users building and utilizing the correspondences. In the conceptual design, I used the circle image to get a front view of the sphere. However, I found that this method only shows the shape of the picture on UI (Figure 24). The front view of the sculpture has a lot of lines representing the different areas of the sculpture. To solve this problem, I extracted the lines from the black and white line drawing as a png that the rest parts are transparent. In the new *image*, I set it to *source image*. Eventually, after adding the lines, I got the ideal front view image of bust.

**Figure 24 Shape of front view image (left) and final view image (right)**

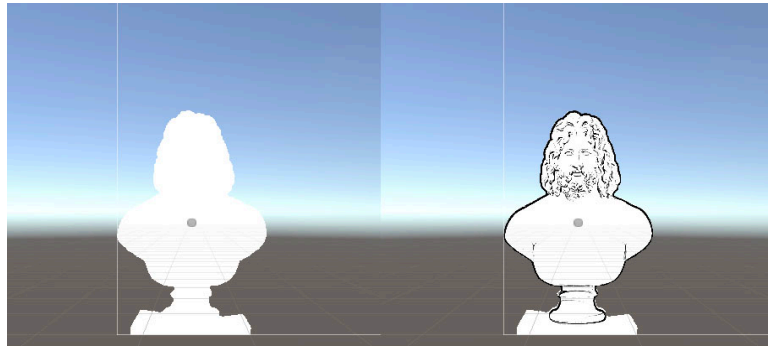

### 3.4.2 Connected to Augmented Reality

#### 3.4.2.1　Image Trager

Image are detected based on natural features that are extracted from the image target and then compared at run time with features in the live camera image. To create a trackable that is accurately detected, Images need to have rich detail, good contrast, avoid repetitive patterns and featureless areas, and formatting must be 8- or 24-bit PNG and JPG formats; less than 2 MB in size; JPGs must be RGB or greyscale (no CMYK). I use this black and white line drawing as the image target (Figure 25).

**Figure 25 Image target**

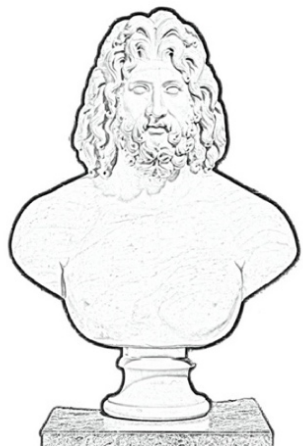

For setting of *ImageTarget*, I place the model vertically on the image target allows the user to interact with the model by simply placing the image target in a horizontal position when running the program (Figure 26).

**Figure 26 Zeus bust on image target in Unity**

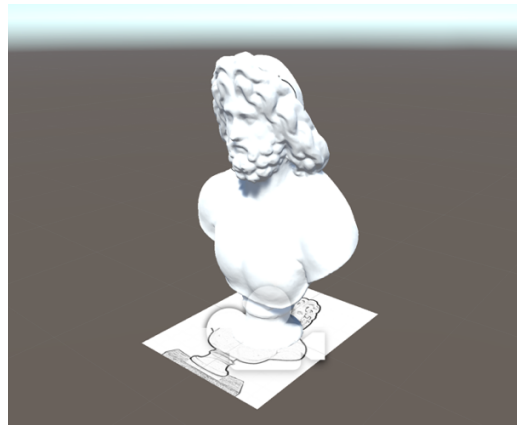

### 3.4.2.2    Augmented Reality Interface

In order to be able to test it on an iPad pro, I set the screen to 1149*843, which will fit the screen. The basic components of Interfaced consist of a Colour Panel, buttons, a black and white line drawing, and a 3D model with the same layout as in the prototype (Figure 27). The user flow of interaction in the AR system is also identical to that of the prototype, except that the model rotation can follow the rotation of the camera angle of view.

**Figure 27 Interface of *Scene0* (left), *Scene1* (middle), and *Scene2* (right)**

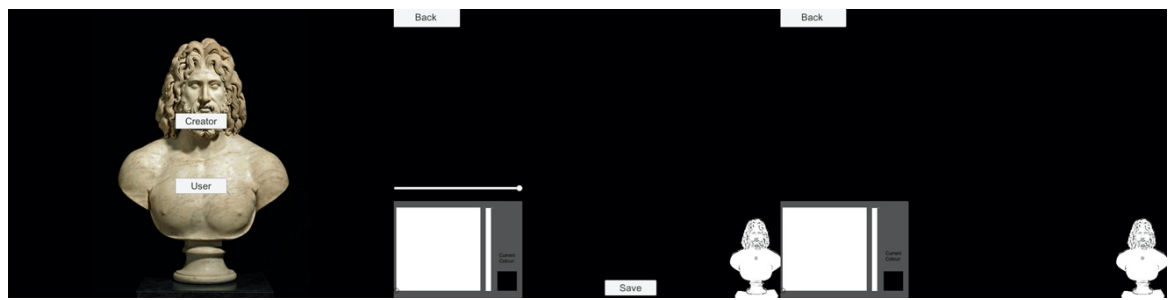

So far, all the setup and build of the prototype is complete. The next step is to install it on the iPad pro for testing.

# 4. Testing

The testing chapter will design the test case as an independent module for each part of system, and each module will be subdivided into tests for each component. A portion of the concept prototype testing including the impact of the number of brushes on the system and detection of distance variable (f value) will be on a laptop, and the AR prototype portion will be on an iPad. The technical specifications of the testing device will be described in detail below.

## 4.1 Testing Devices

This project used Apple MacBook Pro 2020 and Apple iPad Pro 12.9-inch with 2nd generation Apple pencil for testing performance of prototype. The processor of Apple MacBook Pro 2020 is 2.0GHz 10th-generation quad-core Intel Core i5 and 16GB of 3733MHz LPDDR4X onboard memory [41]. The technology specifications of iPad are Apple A12Z Bionic chip. The rear camera with wide 12MP and ultra-wide 10MP, the screen is 2732x2048-pixel resolution at 264 pixels per inch (ppi) with LED-backlit Multi-Touch display and ProMotion technology with refresh rates of up to 120Hz [42].

## 4.2 Conceptual Prototype Testing

### 4.2.1 Brush Sprites

When the creator uses brushes, no matter how many brushes sprites are generated before the user presses the Save button, all sprites are not lost until the system crashes, which means that reaching a certain number of sprites will cause the system running slow. In extreme tests, it is generated more than 6000 sprites of brushes in 4 minutes, the system still runs smoothly (Figure 28), but if more than 3000 brush sprites are generated in a single area, the system will crash. So, it must be saved before that to reduce the cache pressure.

**Figure 28 Debug log for point number in Console window**

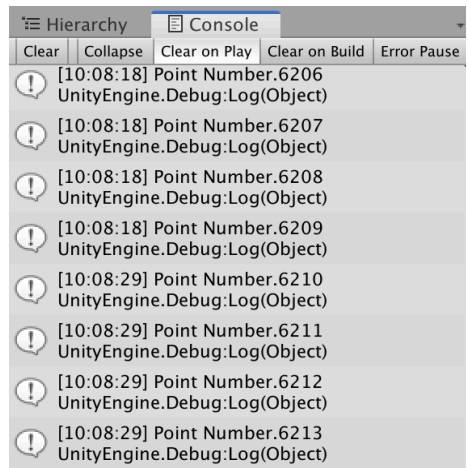

## 4.2.2 Distance Variable

To test distance variable (f value), I chose 11 values from 0.1 to 1000, and the goal of the test is to find numbers that I could just colouring without bias. After test, I find that it is difficult to colouring directly at 0.1 to 1f, although this value range create a more exact match, and allow accurate colouring even two areas are closer together. In 3f to 5f range, it cannot be coloured directly on the matching area every time. It is very easy to colour to the matching areas from 8f. After 50f, if more than one matching area is created close to each other, the colouring area will be bias resulting in colouring errors (Table 3). In the end, I chose 8f as the standard f value for the prototype.

Table 3 Testing f value

| Module | Description | Expected | Actual | Pass/Fail |
|---|---|---|---|---|
| f value | 0.1f | Colouring | cannot | Fail |
| | 0.5f | Colouring | cannot | Fail |
| | 1f | Colouring | hard | Pass |
| | 3f | Colouring | middle | Pass |
| | 5f | Colouring | middle | Pass |
| | 8f | Colouring | easy | Pass |
| | 10f | Colouring | easy | Pass |
| | 20f | Colouring | easy | Pass |
| | 50f | Colouring | bias | Pass |
| | 100f | Colouring | bias | Pass |
| | 1000f | Colouring | bias | Pass |

## 4.3 Augmented Reality Testing

### 4.3.1 Target Registration

Factors that affect registration have been summarised in the table 4. And I printed the image target in black and white on an A4 sheet of paper for following test.

Table 4 Testing image target registration

| Module | Description | Expected | Actual | Pass/Fail |
|---|---|---|---|---|
| Angles of camera | Recognition speed. | Low effect | More parallel the camera is to the target | Pass |
| Distance | Accuracy of | Low effect | Model will | Pass |

| | tracking | | disappear when target loses in screen | |
|---|---|---|---|---|
| System latency | System latency affects registration | Low effect | No significant latency | Pass |
| Variable lighting | Changes in the intensity of the light | Low effect | No significant difference in recognition. | Pass |
| Shadows | Shadows on the image target | Low effect | Recognition latency. | Pass |

## 4.3.2 Interface Interaction

This section tests the interactions in the AR system interface. All components are divided into three modules according to their scene in Unity (Table 5).

**Table 5 Testing interface interaction**

| **Module** | **Component** | **Expected** | **Actual** | **Pass/Fail** |
|---|---|---|---|---|
| **Scene0** | Creator button | Click to entry Creator mode | Entry Creator mode | Pass |
| | User button | Click to entry user mode | Entry user mode | Pass |
| **Scene1** | ColorPanel | Select by saturation | Get colour in saturation | Pass |

|  |  | ColorRGB | Select by hue | Get colour in hue | Pass |
|---|---|---|---|---|---|
|  |  | Brush Size bar | Drag to control size | Dragging to the left is getting smaller. | Pass |
|  |  | Brush | Mark on 3D model | Brush showing on 3D model | Pass |
|  |  |  | Mark on image | Brush showing on image | Pass |
|  |  | Back button | Click to back to Scene0 | Back to Scene0 | Pass |
|  |  | Save button | Click to Save relation | Save relation | Pass |
|  |  | Revoke button | Click to delete unsaved brushes | Unsaved brushes are deleted | Pass |
|  |  | 3D model | Be marked by brush | Marked colour | Pass |
|  |  |  | Drag to rotate | Rotated to any angle | Pass |
|  |  | Image | Be marked by brush | Marked colour | Pass |
| **Scene2** |  | ColorPanel | Select by saturation | Get colour in saturation | Pass |
|  |  | ColorRGB | Select by hue | Get colour in hue | Pass |

| | | | |
|---|---|---|---|
| Brush | Mark on image | Brush showing on image | Pass |
| Back button | Back to Scene0 | Backed to Scene0 | Pass |
| Image | Filling by brush | Filled colour | Pass |
| 3D model | Be marked by relation | Filled colour same as image | Pass |
| | Drag to rotate | Rotate to any angle | Pass |

### 4.3.3 Running Performance

For creating relation, the creator could use colours to mark areas of the bust model and image by and adjusting the size of brush. Then, they could rotate the camera or image target to get a different view of the model (Figure 29). While the program is running, the creator could temporarily remove the iPad's camera from the image target and then to get the bust model by aligning the camera with the image target in any direction without losing the previous marker information.

**Figure 29 Testing marking in AR system**

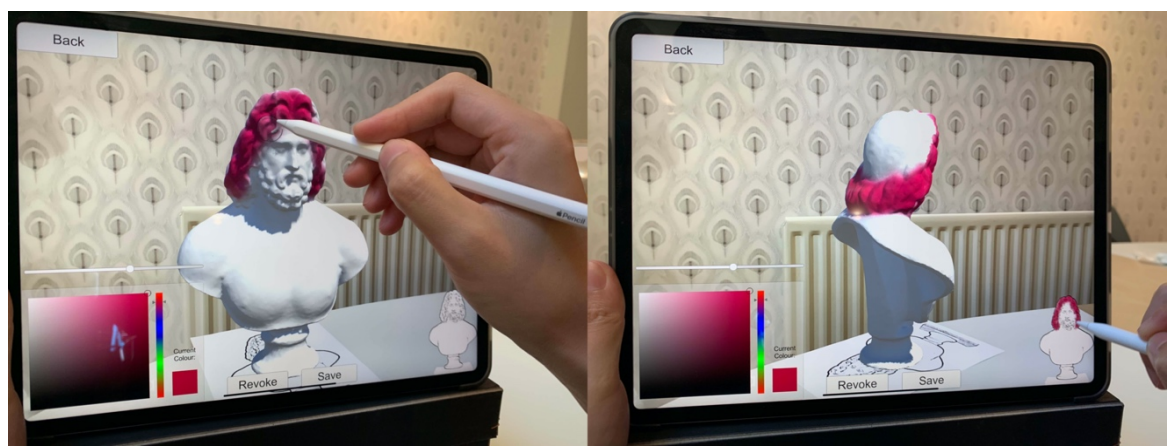

For colouring，standard distance 8f is viable for this statue filling colour and there is no bias in the colouring process (Figure 30). The end users can rotate the camera around the image target while the system tracking it to see the coloured bust at any angle. If the model is lost from the screen, they can get the coloured bust by putting camera refacing to image target again.

**Figure 30 Testing filling colour in AR system**

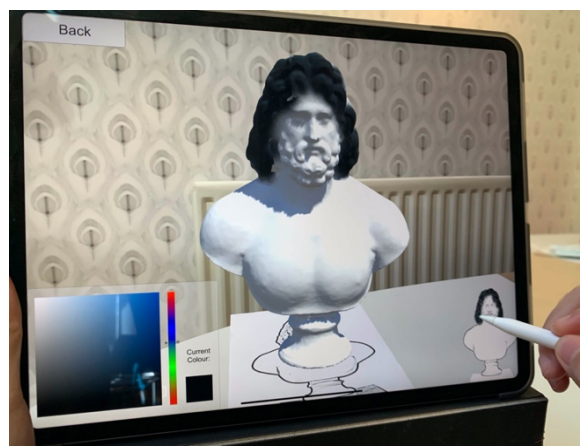

A discussion of the interpretations and implications and limitations of the test is presented in the next chapter.

# 5. Discussion and Future Work

## 5.1 Interpretations and Implications

3D painting had already begun to develop programs to paint directly on the model and to change different brushes and colours to simulate the effects of hand painting. Hanrahan and Haeberli believe that 3D painting should be programmed to allow the user should directly see the interactive results [12], which becomes the basis of concept the development of this prototype. After that, the use of haptic screens in development of 3D painting [22], and it also became the equipment support for AR colouring project. Appreciate to the development of UV mapping, it has become very easy to use texture maps to colour 3D models [15]. Now, this method is widely used in AR colouring projects, whether it uses to diffuse texture [10, 33] or uses the both copy and diffusion method [2], it is difficult to develop more complex models to get a result due to the limitations of texture maps. In addition, these projects only consider the end user experience, and ignore the process of establishing matching. So, the initial idea of this project is a challenge to implement colouring the sculpture through corresponding area from 2D to 3D with augmented reality. In the following comparison and summary, it is found that the use of render texture in 3D model proved the feasibility of colouring, and it is applied to the system for establishing corresponding relations by colouring the model and the image with both render textures.

Future projects should take into consideration the findings of this prototype because the most important contribution of the prototype is to take creator into account the user groups with a wilder perspective to build a system that is easier to establish 2D to 3D relations, which means that the workflow for AR colouring might be reduced and the system is more open to allow anyone to establish matching relations. More and more complex models will be added to the AR colouring system, not only for public entertainment and educational purposes, but also to assist restorers in important areas such as museum restoration.

## 5.2 Limitations

As a result of the tests above, there are still some limitations in prototype. First, in the interface design, the model's front view area is too small which makes it difficult to create areas and fills using brushes, the size of the brushes needs to be adjusted more appropriately to the interface design. Second, in the process of building relations, although losing the model on the screen will not lose the relation information, holding the iPad for a long time will be very tiring for the users, the interactive flow and the presentation of the model need to be iteratively optimized. Third, it is not friendly that the end user can't change the colour again after colouring, which should be a function to be added in the future.

## 5.3 Future Work

In a wide perspective, there are two areas of work that could be undertaken in the future, including building database and connecting 3D print. The database is a system for organizing, storing and managing data according to data structures. One of the ways in which future applications will work is to connect to the database, use this connection to store the data generated by the application, and then colour the model using the relations established by the creator within the database once the user is connected to the network.

3D printing is a type of rapid prototyping technology in which a computer-designed three-dimensional digital model is broken down into several layers of flat slices, and then a 3D printer stacks the different bondable materials layer by layer in a slicing pattern to eventually build up a complete object. The ultimate future presentation of this project could be a 3D printed object as the final output, with the foreseeable future being the use of coloured texture information to generate new model data that can be exported as a solid coloured model via a pre-programmed 3D printing device. In the function. if a more advanced method is used to improve the accuracy of the matching relations, adding texture maps to compensate for model detail, and linking to more advanced large-scale 3D printers working in conjunction, which could provide a cheaper and more convenient option for the museum's restoration efforts and reconnecting to the real world. There has been a case of painted models being 3D printed, but there are still some 3D printing material limitations that may cause the colours to darken or become fragile [42].

# 6. Conclusion

This project implements three different methods to 2D to 3D region matching in the conceptual design. The first method is discovered during the modelling process and is used by most of the current programs for simple models, but the actual results from the experimental Caligula model are not perfect and take a lot of time. The second method has been optimized from the first, by adding multiple materials and grouping the faces of the high poly model according to the model structure that creator want to divide. This is the result of using advanced Blender software, but it is still not friendly for establishing a large number of 2D-3D correspondences, and the fine procedure depends on the model itself. The third method is to use marker colours to establish correspondence between render textures of image and 3D model. Although it temporarily cannot to add the original texture map and change the filled colour, this method can handle any model with UV mapping properly, which reducing the large amount of workload of modelling for creator to establish correspondence. Such correspondences support many-to-many relations, and there is no limit to the number of them.

In order to achieve the aim of simpler mapping creation, I chose to the render texture mapping concept and eventually tested the Zeus bust model in the AR environment. The test results showed that the idea of pointing to the UV map, marking on render textures, and saving relations involved in concept of render texture mapping meet the project aim, and the tests discover and then solve some of problems that appeared in the prototype including the number of brushes sprites, distance detection, setting of model front view image. Three major limitations and future work are: 1. image size needs to be adjusted to fit the brush size and layout; 2. tiredness during utilizing requires iterative optimization of the interaction process using additional user data and device selection; 3. the end user cannot yet change the colour after filling in the region. The broad future expectation is that the program can utilise the database to provide end users with mapped data produced by creator, even each user can create and use the correspondences data as both creator and end user, and finally, by connecting to advanced 3D printers to produce physical models.